\documentclass[12pt]{iopart}

\usepackage{bm}
\usepackage{graphicx}
\usepackage{color}
\usepackage{subfigure}
\usepackage{hyperref}
\usepackage{latexsym}
\usepackage{amsthm}
\usepackage{amssymb}
\usepackage{braket}
\usepackage{cite}
\usepackage[rgb]{xcolor}
\usepackage{hyperref}

\usepackage{bbold}
\usepackage{bbm}
\usepackage{verbatim}
\usepackage{multirow}
\usepackage[english]{babel}
\usepackage{comment}
\usepackage{bm}
\DeclareGraphicsExtensions{.jpg,.pdf, .mps, .png, .eps, .ps, .EPS,.gif}

\begin{document}

\def\be{\begin{equation}}
\def\ee{\end{equation}}

\def\bc{\begin{center}}
\def\ec{\end{center}}
\def\bea{\begin{eqnarray}}
\def\eea{\end{eqnarray}}
\newcommand{\avg}[1]{\langle{#1}\rangle}
\newcommand{\Avg}[1]{\left\langle{#1}\right\rangle}

\def\ie{\textit{i.e.}}
\def\etal{\textit{et al.}}
\def\m{\vec{m}}
\def\G{\mathcal{G}}

\newcommand{\davide}[1]{{\bf\color{blue}#1}}
\newcommand{\gin}[1]{{\bf\color{green}#1}}

\newcommand{\change}[1]{{\color{red}{#1}}}

\title[Classical information theory of networks]{Classical information theory of networks}

\author{Filippo Radicchi}
\address{Center for Complex Networks and Systems Research, Luddy School
  of Informatics, Computing, and Engineering, Indiana University, Bloomington,
  IN 47408}
  
\author{Dmitri Krioukov, Harrison Hartle}
\address{Department of Physics, Department of Mathematics,
Department of Electrical \& Computer Engineering, Northeastern University, Boston, MA 02115, USA}

\author{Ginestra Bianconi}

\address{School of Mathematical Sciences, Queen Mary University of London, London, E1 4NS, United Kingdom\\
Alan Turing Institute, The British Library, London, United Kingdom}
\ead{ginestra.bianconi@gmail.com}
\vspace{10pt}
\begin{indented}
\item[]
\end{indented}

\begin{abstract}
Existing information-theoretic frameworks based on maximum entropy network ensembles are not  able to explain the emergence of heterogeneity in complex networks.
Here, we fill this gap of knowledge by developing a classical framework for networks based on finding an optimal trade-off between the information content of a compressed representation of the ensemble and the information content of the actual network ensemble.
In this way not only we introduce a novel classical network ensemble satisfying a set of soft constraints but we are also able to  calculate the optimal distribution of the constraints.
We show that for the classical network ensemble in which the only constraints are the expected degrees a power-law degree distribution is optimal.
Also, we study spatially embedded networks finding that the interactions between nodes naturally lead to non-uniform spread of nodes in the space, with pairs of nodes at a given distance not necessarily obeying a power-law distribution.
The pertinent features of real-world air transportation networks are well described by the proposed framework.
\end{abstract}

%
%
%
%
%

\section*{Introduction}

The principle of maximum entropy states that the unique probability
distribution, encoding all the information available about a
system but not any other information, is the one with largest information
entropy~\cite{jaynes1957information}. Available information about the
system corresponds to constraints under which entropy is maximized.
The principle of maximum entropy has found applications in many different disciplines,
including physics~\cite{presse2013principles}, computer
science~\cite{ratnaparkhi1997simple}, geography~\cite{WILSON1967253},
finance \cite{ravi}, molecular biology~\cite{chen},
neuroscience~\cite{tang2008maximum},  learning \cite{tishby2000information}, deep learning
\cite{song2018resolution}, etc.

Powerful information-theoretical frameworks that extend and generalize
maximum entropy principles { by making use of operation such as
  compression or erasure of information} have been recently
proposed.
{A paradigmatic example is the}  information bottleneck
principle~\cite{tishby2000information}. {The principle
  allows to optimally learning a given output from an input signal,
and the optimization relies on } finding the  best trade-off between accuracy of the prediction 
and effectiveness of the compression.  Another notable  {
  example of this type of theoretical frameworks is}
  the study of computation
  and the investigation of the entropic cost of bit erasure \cite{wolpert2019stochastic}.

Applications of the maximum principle  
can be found also in network science~\cite{cimini2019statistical,park2004statistical,bianconi2009,
anand2009,anand2010,bianconi2007statistical,peixoto2012entropy,horvat2015reducing,sagarra2013statistical}, where
the maximum entropy argument is  applied to the distribution
of probabilities~$P(G)$ of observing a given graph~$G$ of finite size
$N$ in an ensemble of random graphs.
Different  entropy-maximization constraints lead to
different network models. For example, if the constraints are soft,
i.e., if they deal with expected values of network properties, then
$P(G)$ is a Gibbs-like distribution corresponding to
ERGMs~\cite{park2004statistical,holland1981exponential}.

 {
This approach can be used to model networks with  
heterogeneities, e.g., in node degrees~\cite{barabasi1999emergence, voitalov2019scale}, edge weights~\cite{barrat2004architecture}, and community
sizes~\cite{radicchi2004defining, lancichinetti2008benchmark}.
However, an important shortcoming of this approach is that it  
 cannot explain why these heterogeneities can  be found so ubiquitously in real networks.}
Indeed, current maximum entropy approaches can only generate the
least biased network ensembles with given expected degree sequence,
but they cannot be used to explain or justify why in many cases
we observe heterogeneous degree sequences.
Similarly for spatially embedded networks, current maximum entropy approaches
can be used to provide ensembles of spatial networks for a given
distribution of nodes in the space, 
but  {they} cannot be used to draw any conclusion on the expected spatial
distribution of the nodes in the network. 
Therefore if we we want to infer the positions of the nodes in
network 
embedding we do not have information theory guidelines on how to
choose the prior on the spatial distribution of the nodes.

In the present paper, we address this fundamental shortcoming of
 {current information-theoretical approaches to the study of networks.}
Specifically,  we derive
 {a novel}
framework that is
able to predict the optimal degree distribution 
and the optimal spatial distribution of nodes in space. Both
distributions turn out to be heterogeneous, thus
providing a principled explanation of the origin of 
heterogenities in complex networks. Our approach is based on finding the best 
compressed representation of a network ensemble, given the content of
information conveyed by
 { the} ensemble. 
We consider network ensembles
 {where}
any pair of nodes is associated with a set of hidden variables
 {obeying}
an arbitrary distribution, e.g., arbitrary degree distribution or arbitrary distribution of
distances between pair of nodes, 
expressed in general as $P_{{\mathcal V}}({\bf x})$.
We measure the
information content of the network ensemble and of its compressed
network ensemble representation in terms the corresponding entropies
$S$  and  $H$, respectively. 
Finally we propose to find the optimal hidden variable distribution
$P_{{\mathcal V}}^{\star}({\bf x})$, e.g., degree distribution or
spatial distribution of distance between pair of nodes,
by maximizing 
\bea
P_{{\mathcal V}}^{\star}({\bf x})=\arg\max_{{P_{\mathcal V}}({\bf x})} [H-\lambda S]
\eea
under the constraints that the network contains a given number 
of nodes and links,  { and that the entropy of the network ensemble is given,
i.e., $S=S^{\star}$.}
As explained in the main text and in the Appendices, this principle is
solidly rooted in information theory \cite{mackay2003information} as
the classical network ensemble and its compressed representation can
be seen respectively as the input and output of a communication
channel. Therefore  the definition of the
optimal hidden
variable distribution
can be 
 {interpreted as}
the
optimal input distribution
of a communication channel in information theory. 


We believe that our results not only provide an
information-theoretical explanation for the emergence of heterogeneous properties
in complex networks, but also open a  promising perspective for
devising a new generation of inference methods
 {for finding optimal network embeddings.}

\section*{Results}

\subsection*{Classical network ensembles\\}

 {The simplest examples of maximum entropy ensemble are the $\mathcal{G}(N,p)$ and $\mathcal{G}(N,L)$ ensembles obtained by enforcing a constrain on the expected and the actual total number of links, respectively~\cite{erdds1959random,bollobas}.
In network theory these ensembles can be respectively generalized to canonical and microcanonical network  ensembles enforcing a set of soft and hard constraints\cite{bianconi2009,anand2009,anand2010} which are not in general equivalent \cite{anand2009,anand2010,garlaschelliprl}.
A major example of canonical network ensemble is the exponential
random graph mode (ERGM) enforcing a  given expected degree sequence~\cite{park2004statistical} whose  conjugated microcanonical ensemble is the configuration model \cite{molloy1995critical,anand2009,anand2010}. }

In all the examples above, the maximum entropy principle is {\it de facto} applied to network adjacency matrices~$A$ 
whose elements are understood as sets of edge variables 
correlated by the imposed constraints. Calculations generally
lead to the derivation of the probability $\pi_{ij} = P(A_{ij} = 1)$ 
for the pair of nodes $i$ and $j$ to 
be connected. If networks are undirected, then $A_{ij} = A_{ji}$ and $\pi_{ij} = \pi_{ji}$. This approach is very similar to the one used in  {quantum} statistical mechanics to describe systems of noninteracting particles whose role is played by network edges, while particle states are enumerated by node pairs~$(i,j)$~\cite{park2004statistical,bianconi2009}. 
 {In fact the adjacency matrix element $A_{ij}$ indicating the number of links between a pair of nodes $(i,j)$ corresponds to the ``occupation number" in quantum statistical mechanics.} 
Indeed in binary networks, where $A_{ij}$ is either $0$ or $1$, $\pi_{ij}$ takes the Fermi-Dirac form; if multiple edges are allowed between the same pair of nodes, then the system is described by the Bose-Einstein statistics~\cite{garlaschelli2009generalized}.

Here we take advantage of the principle of maximum entropy
in a classical
way. Instead of dealing with all elements
of the adjacency matrix
(corresponding to the occupation numbers of quantum statistical mechanics)
we look directly at network edges (which corresponds to particle states). Therefore a given network $G$ of $N$ nodes is identified by its edge list  $\left\{\vec{\ell}^{[n]}\right\}$ with $n\in\{1,2\ldots, L\}$ where each link $\vec{\ell}^{[n]}$ is described by an ordered pair of node labels  $\vec{\ell}^{[n]}=\left(\ell_1^{[n]},\ell_2^{[n]}\right)$, i.e.,  $\ell_1^{[n]}$  indicates the label of the node $i\in\{1,2,\ldots, N\}$ attached to the first end of the link $n$ and similarly $\ell_2^{[n]}$ indicates the label of the node attached to the second end of the link $n$ (see Appendices for details).

We assume that the ends of each link are drawn independently from the probability distribution
$P(\vec{\ell})$. Therefore   $P(\vec{\ell})$ indicates the probability that,
by picking a 
random edge, nodes $\ell_1$ 
and $\ell_2$ are found at its ends. 
The Shannon entropy $S$ of this ensemble is given by 
\bea
S=-L\sum_{\vec{\ell}}P(\vec{\ell})\ln P(\vec{\ell}).
\label{eq:classical_entropy0}
\eea
$S$ is named the
classical entropy
and
quantifies the information content  associated 
with all edges in the network.   {If we indicate with $\avg{k}$ average degree of the network,} Eq.(\ref{eq:classical_entropy0})  indicates that the entropy $S$ is given by the sum of $L
=\avg{k}N/2$
identical terms corresponding to the entropy associated with the typical
number of  ways in which we can choose  two nodes ($i,j$) to be
connected by a single link.

The distribution $P(\vec{\ell})$ that describes
the ensemble is then found using the maximum entropy principle. 
Different constraints in 
the entropy maximization problem lead to 
different distributions $P(\vec{\ell})$. Since the marginal probabilities in this
ensemble are exponential, we refer to it as
the classical network ensemble, 
differentiating it from previously explored maximum entropy
ensembles where the marginals obey quantum statistics~\cite{park2004statistical}.
We note that the framework we consider here allows for multiedges and tadpoles
as in similar approaches~\cite{WILSON1967253,sagarra2013statistical}.
This makes all edges uncorrelated variables, allowing for greater
simplicity and flexibility.

\subsection*{Classical network ensemble with  expected degrees}
As the first very basic example  of  classical network ensemble,
we consider the ensemble in which we constrain  {expected values
  of node degrees.} 
That is, we require that the 
probability to find node $i$ at one of the ends
of a randomly chosen link is $k_i/L$,
\bea
\sum_{\vec{\ell}} P(\vec{\ell}) \, \left[ \mathbbm{1} (\ell_1 =
  i)+\mathbbm{1}(\ell_2 = i)\right]=\frac{k_i}{L}\;,
\label{eq:constraint_cm}
\eea
where $\{k_i\}$ is any given degree sequence, $L$ is a fixed number of links
in the network, which is assumed to be consistent with $k_i$s via $2L=\sum_ik_i$,
and where $\mathbbm{1} (x=y)$ is the indicator function:
$\mathbbm{1} (x=y) = 1$ if $x=y$ and $\mathbbm{1} (x=y) = 0$ otherwise.
The constraint in Eq.~(\ref{eq:constraint_cm}) is required to hold for all
nodes $i = 1,\ldots, N$.

The maximum entropy distribution $P(\vec{\ell})$ is found by maximizing the functional
\bea
\hspace{-20mm}{\mathcal G}&=&S-\mu L\left[\sum_{\vec{\ell}}P(\vec{\ell})-1\right] 
=-L \sum_{i=1}^N \psi_i\left[\sum_{\vec{\ell}}P(\vec{\ell})\left[
    \mathbbm{1}(\ell_1 = i)+\mathbbm{1} (\ell_2 = i)\right]-\frac{k_i}{L}\right].
\label{eq:max_entropy_cm}
\eea
where we have introduced the  Lagrange multipliers $\psi_i$ and $\mu$ associated with the constraint in Eq.~(\ref{eq:constraint_cm}) and the normalization of~$P(\vec{\ell})$, respectively. 
The solution of this maximization problem leads to the expression for
the probability $\pi_{ij}$ that a given link connects node $i$ at one
end to node $j$ at the other end, that is
\bea
\pi_{ij} = P(\ell_1 = i, \ell_2 = j) = e^{-\mu}e^{-\psi_i-\psi_j}
\; , \eea
where the Lagrange multipliers $\psi_i$ and $\mu$ are the solutions of the constraint equations
\bea 
\frac{k_i}{L}=2\frac{k_i}{\avg{k}N}=2
e^{-\mu}e^{-\psi_i}\sum_{j=1}^N e^{-\psi_j}\; . \eea

Therefore,
$e^{-\psi_i}=k_i$ and $e^{\mu}=(\avg{k}N)^2$, from which we obtain 
\bea
\pi_{ij}=\frac{k_i \, k_j}{(\avg{k}N)^2} \; .
\label{eq:probability_cmp}
\eea
Notice that $\pi_{ij}$ is the probability that a link connects node $i$ at the first end and node $j$ at the second end, therefore the $\pi_{ij}$ is a distribution and   obeys the 
normalization condition $\sum_{ij}\pi_{ij}=1$.
Since there are $L=\avg{k}N/2$ links in the network, and two nodes are connected if there is a link attached to the two ends in any possible order, the average number of links that connect node $i$ to node $j$ is given by 
\bea
\langle A_{ij} \rangle =2L\pi_{ij}=\frac{k_ik_j}{\avg{k}N}.
\label{eq:probability_cm}
\eea
This is the average number of links  
between nodes of degrees $k_i$ and $k_j$ in uncorrelated random networks~\cite{molloy1995critical}.
Equation~(\ref{eq:probability_cm}) is the starting point of many calculations in network
science that use the uncorrelated random networks as a null model.
A popular example is the modularity function used in community detection~\cite{newman2004finding}.
The derivation above provides a theoretical ground for such an interpretation of the model.

We now turn to more sophisticated outcomes of the considered framework. 
In  our classical network ensemble, the degree distribution $P(k)$ is an input parameter that we can set to whatever we wish.
Among all possible choices of the degree distribution, which one
corresponds to maximal randomness?

To answer this question, we note that we can express the 
classical network entropy $S$ in terms of the degree distribution
$P(k)$ as
\bea
S&=&-L\sum_{ij}\pi_{ij}\ln \pi_{ij} =  \avg{k}N \left[\ln (\avg{k}N)- \sum_k \frac{kP(k)}{\avg{k}} \ln k\right]\;.
\label{eq:classical_entropy}
\eea
The entropy~$S$ quantifies the amount of information encoded in the classical network  ensemble with $N$ nodes, $L$ edges, and degree distribution $P(k)$.
Any given $P(k)$ uniquely determines the value of~$S$ via Eq.~(\ref{eq:classical_entropy}), yet the same value of~$S$ may correspond to different~$P(k)$s.

\subsection*{Information theory framework\\}

Here we describe our theoretical framework to predict the optimal degree distribution in 
terms of a standard information-theoretic problem~\cite{mackay2003information}.
A network instance is a ``message.''  Specifically, a message
consists of $L$  two-letter words, each representing a link
$\vec{\ell} = (i, j)$.
Letters are node labels, so that the alphabet is given by $N$ distinct
symbols. We assume that messages are generated
by picking random pairs of nodes according to  the probability $\pi_{ij}$. 
For the classical network ensemble enforcing expected degrees the probability $\pi_{ij}$ is only dependent on the degrees of the nodes, i.e.,
\bea \pi_{ij}=\pi_{k_i,k_j}\eea
with 
\bea
\pi_{k,k'}=\frac{kk'}{(\avg{k}N)^2},\eea
indicating  the probability
that two nodes of degree $k$ and $k'$ are connected to one or the other end node of a link in the classical network ensemble.
This is our source of  messages. If we change degree sequence, then we
have a different source of messages.
The entropy $S$ defined in Eq.~(\ref{eq:classical_entropy})
is the entropy of the source. In our specific setting, $S$ turns out
to be a function of the degree distribution $P(k)$ only, not of the
specific degree sequence $\{k_1, \ldots, k_N\}$.
Thus, if we change the degree
distribution $P(k)$, then we change the source of messages.

Once generated, messages are compressed using a lossy compression channel.
The choice of the channel is naturally suggested by the classical network ensemble under consideration.
Since for  the classical network ensemble enforcing expected degrees the probability $\pi_{ij}=\pi_{k_i,k_j}$ only depends on the two node degrees, we use the channel where the link labels $(i,j)$ are replaced
with the link label of the pair of degrees $(k,k')$  of the two linked nodes.
Please note that the messages are still the same as those generated by the source. However, many of them are not longer distinguishable after the application of the channel.
Specifically the channel is erasing information about the actual identity of the linked nodes and is retaining only the information about their degrees.

\begin{figure*}[!htbp]
\begin{center}
\includegraphics[width=0.7\textwidth]{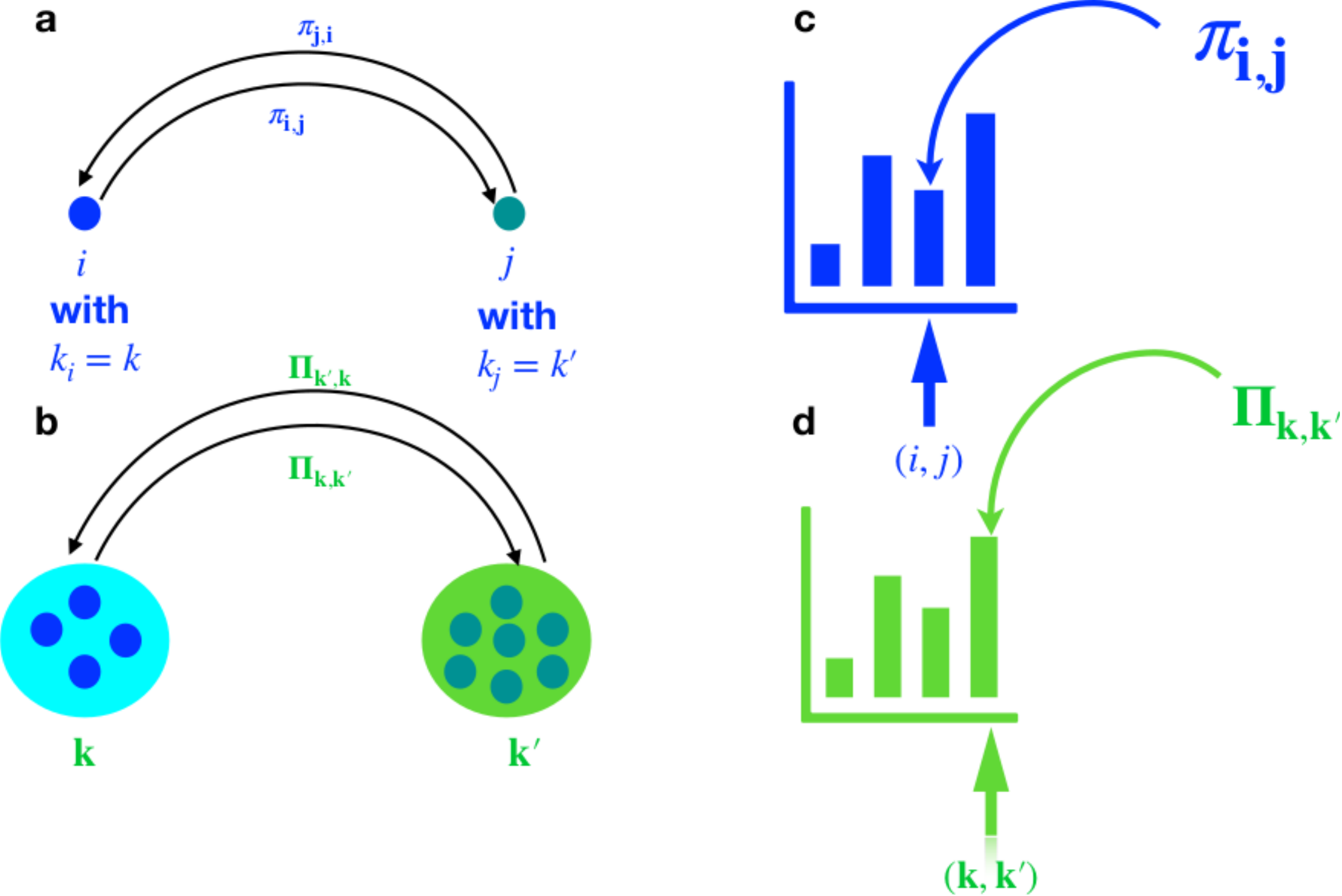}
\caption{{\bf Classical network ensemble with expected degree and its
    compressed representation} (a) Schematic representation of the
  classical network ensemble with expected degrees.   (b) Schematic representation of
  the compressed classical network ensemble. 
  (c)  The classical network ensemble constituting our source is
  defined by the probability $\pi_{ij}$ that a link is attached at one
  end to node $i$ and at the other end to node $j$. This ensemble
  constitutes the source of messages of our channel.(d) The compressed
  classical network ensemble constituting a lossy compression of the classical network ensemble is  defined by
  the probability $\Pi_{kk^{\prime}}$ that a link is attached at one
  end to node of degree $k$ and at the other end to a node  of degree
  $k^{\prime}$. This ensemble is
  the output of our channel that performs a lossy compression of the source of messages.
}
\label{fig:diagram1}
\end{center}
\end{figure*}
The output of the channel  corresponds to a  coarse-grained network ensemble (see Fig. \ref{fig:diagram1}), where all nodes with the same degree class are indistinguishable and they form  a super node in the coarse-grained description.
The network ensemble can be used to compress the information of the
original network retaining only the information regarding the degree
of the linked nodes. Clearly, in this ensemble we observe the
same expected number of links $L_{k,k'}$ between nodes of degree $k$
and nodes of degree $k'$ as in the classical network ensemble. If we
indicate with $N_k$ the number of nodes in degree class $k$ , it is easy to show that
$L_{k,k'}=L\pi_{k,k'}N_kN_{k'}$ and that 
$\sum_{k,k'}L_{k,k'}=L$.
Every link of the coarse-grained ensemble has probability $\Pi_{k,k'}$ to connect super-nodes corresponding to  degree classes $k$ and $k'$ where 
\bea
\Pi_{k,k'}=\frac{L_{k,k'}}{L}=\frac{kk'P(k)P(k')}{\avg{k}^2}.\eea
This compressed ensemble is on its own a classical network ensemble, therefore its entropy $H$  is
\bea
H =-L\sum_{k,k'} \Pi_{k,k'}\ln \Pi_{k ,k'}=- \avg{k}N \sum_{k}\frac{kP(k)}{\avg{k}}\ln \left(\frac{kP(k)}{\avg{k}}\right).
\label{eq:entropy_links}
\eea
We have two representations of the network ensemble at the node level and at the compressed level whose   information content is quantified respectively by the $S$-entropy and the $H$-entropy.
Note that the different notation is only introduced to distinguish between the entropy of the original ensemble and the entropy of its compressed version. However, the entropy $H$ is nothing else that the entropy of a classical network ensemble whose nodes are degree classes.
Given that our channel is only erasing information, we have the interesting results that the entropy $H$ of the output of the channel  is equal to the mutual information between the input and the output of the channel  (see Appendices) and represents a metric of effectiveness of the channel: the
higher its value, the more effective is the channel in transmitting the information produced by the source.

In summary, we  have potentially many sources of messages given by classical network ensemble with different $P(k)$s, but we have
one given channel  prescribed by our coarse-grained procedure of the network.

The maximization problem that we solve consists in determining the best distribution of hidden variables that maximizes the capacity of our channel for given value of the entropy $S$ of the source. Therefore  we maximize $H$  for fixed value of $S$. The constraint on the entropy $S$ is imposed as we do not
want to compare the performance of the
channel over arbitrary sources of messages,
but only over sources with similar level of information.

The
optimal degree distribution
that will allow the best reconstruction of the original network ensemble given only the knowledge of its compressed representation is given by 
\bea
P^{\star}(k)=\arg\max_{P(k)} [H-\lambda S] \; ,
\eea
where the optimization is performed under the constraints the network contains a given number 
of nodes and links  and that $S=S^{\star}$.

We stress that our problem is formulated as essentially an  optimization of the capacity of the channel aimed at finding the optimal distribution of hidden variables for any fixed value of the entropy $S=S^{\star}$ (see Appendix for more details about the oretical framework).

\subsection*{Optimal degree distribution\\}
We now show how our theoretical framework can allow us to predict the optimal degree distribution of  the classical network ensemble with  expected degrees.
We impose the constraint $S = S^\star$, where $S^\star$ is a given positive real number,  i.e., we consider different network ensembles  that have the same information content or ``explicative power'' at the node level.
To find the typical degree distribution $P(k)$ under this constraint, we maximize the randomness of the coarse-grained model quantified by the $H$-entropy.
Clearly, $P(k)$ must also satisfy the constraints
$\sum_{k}kP(k)=\avg{k}$ and $\sum_k  P(k)=1$. Combining all together,
we thus have to maximize the functional
\bea
{\mathcal F}&=&H-\lambda \left[S-S^{\star}\right]-\mu N\left[\sum
  kP(k)-\avg{k}\right]-\nu  N\left[\sum_k P(k)-1\right]\;,
\label{eq:maximum_entropy_pl}
\eea
from which we obtain
\bea
P^{\star}(k)= \avg{k}e^{-(\mu+1)} \, e^{-\nu\frac{\avg{k}}{k}} \, k^{-(\lambda+1)}.
\label{eq:pl}
\eea
The Lagrange multipliers $\lambda$, $\mu$, and $\nu$ are determined by the imposed constraints and they always exist as long as $\lambda>1$

\begin{figure*}[!htbp]
\begin{center}
\includegraphics[width=0.95\columnwidth]{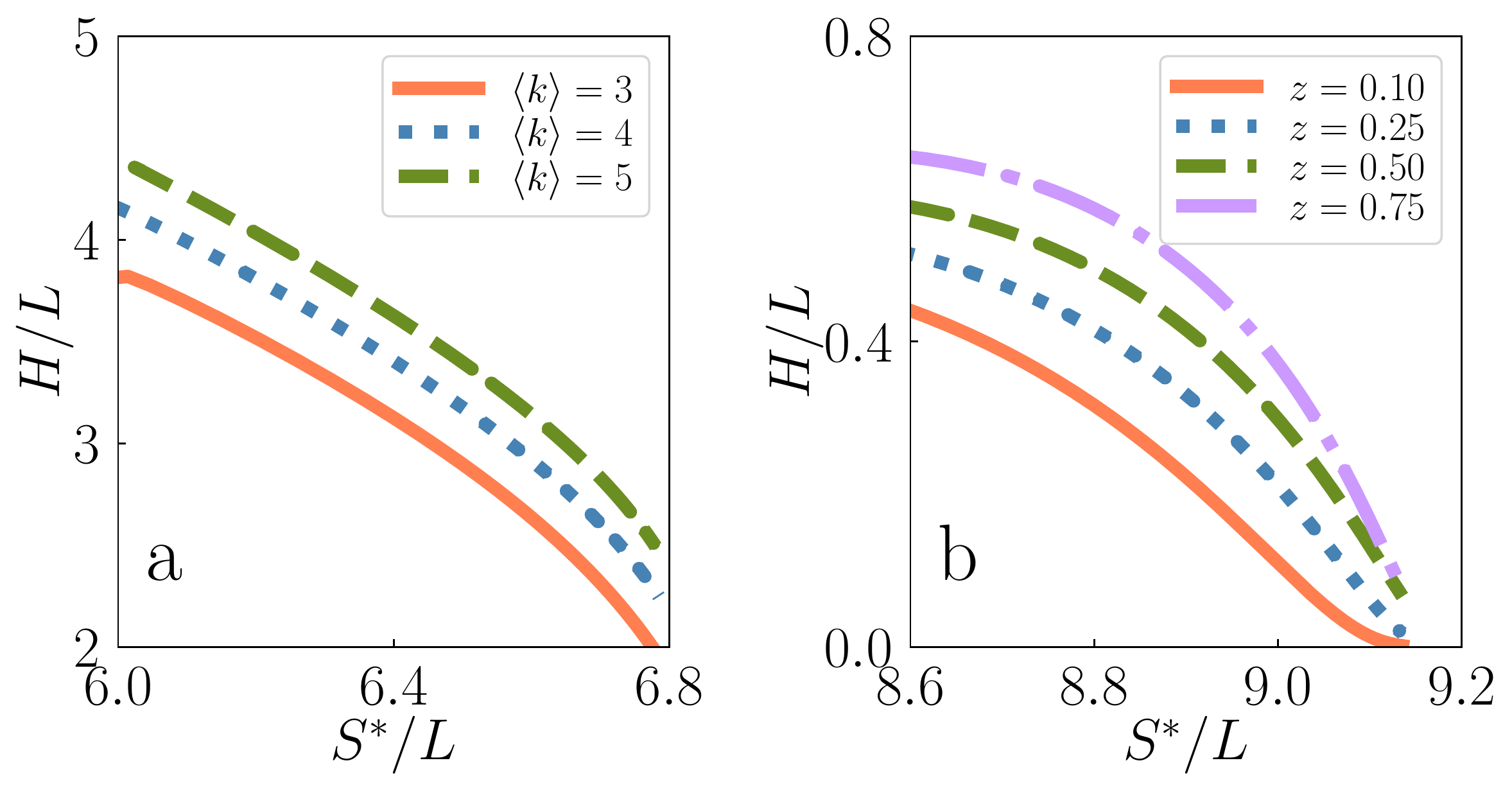}
\caption{{\bf Entropy $H$ as a function of~$S^{\star}$.} (a)~The $H$-entropy in Eq.~(\ref{eq:entropy_links}) is
  evaluated for the degree distribution~$P(k)$ that maximizes
  the functional $\mathcal{F}$ in
  Eq.~(\ref{eq:maximum_entropy_pl}). We consider
  $N=10^4$, and different values of the $S$-entropy constraint
  $S^{\star}$ and
  the average degree $\avg{k}$.
  (b)~Same as in panel~(a), but for the
   spatial ensemble $H$~(\ref{eq:spatial_H}) and $\mathcal{F}$~(\ref{eq:spatial_F})
  with the power-law linking probability
  $f(\delta)=\delta^{-\alpha}/z$. We consider $\alpha=3$ and different
  values of~$z$.
  }
\label{fig:SH}
\end{center}
\end{figure*}

Equation~(\ref{eq:pl}) shows that the optimal degree distribution $P^{\star}(k)$ with a given value of the classical entropy in Eq.~(\ref{eq:classical_entropy}) is a power law.
To be precise, the power-law decay holds for large degrees~$k$,
while in the low-$k$ region there is an exponential cutoff that affects the mean of the distribution.
In Figure~$\ref{fig:SH}$(a) we show the entropy~$H$ as a function of~$S^{\star}$ for different values of the average degree $\avg{k}$.
The lower the~$S^{\star}$, and consequently the lower the power-law
exponent~$\lambda$, the higher the entropy $H$. This is because even
though the number of networks with a given degree sequence decreases
as $S^{\star}$ and $\lambda$ go down, the number of  ways to split $L$
links into classes of links connecting nodes of degrees $k$ and $k'$
increases. Therefore this result highlights the entropic benefit to
have networks with broad (i.e., low $\lambda$ values) degree
distributions
that correspond to low values of the $S$-entropy but to high values of the
$H$-entropy. 
Interestingly, the same result could be obtained by maximizing the
randomness of the classical network ensemble, and therefore optimizing
the $S$-entropy while keeping fixed the informative power of its compressed description, i.e., the $H$-entropy.

\subsection*{Classical information theory of spatial networks\\}

\begin{figure*}[!htbp]
\begin{center}

\includegraphics[width=0.7\textwidth]{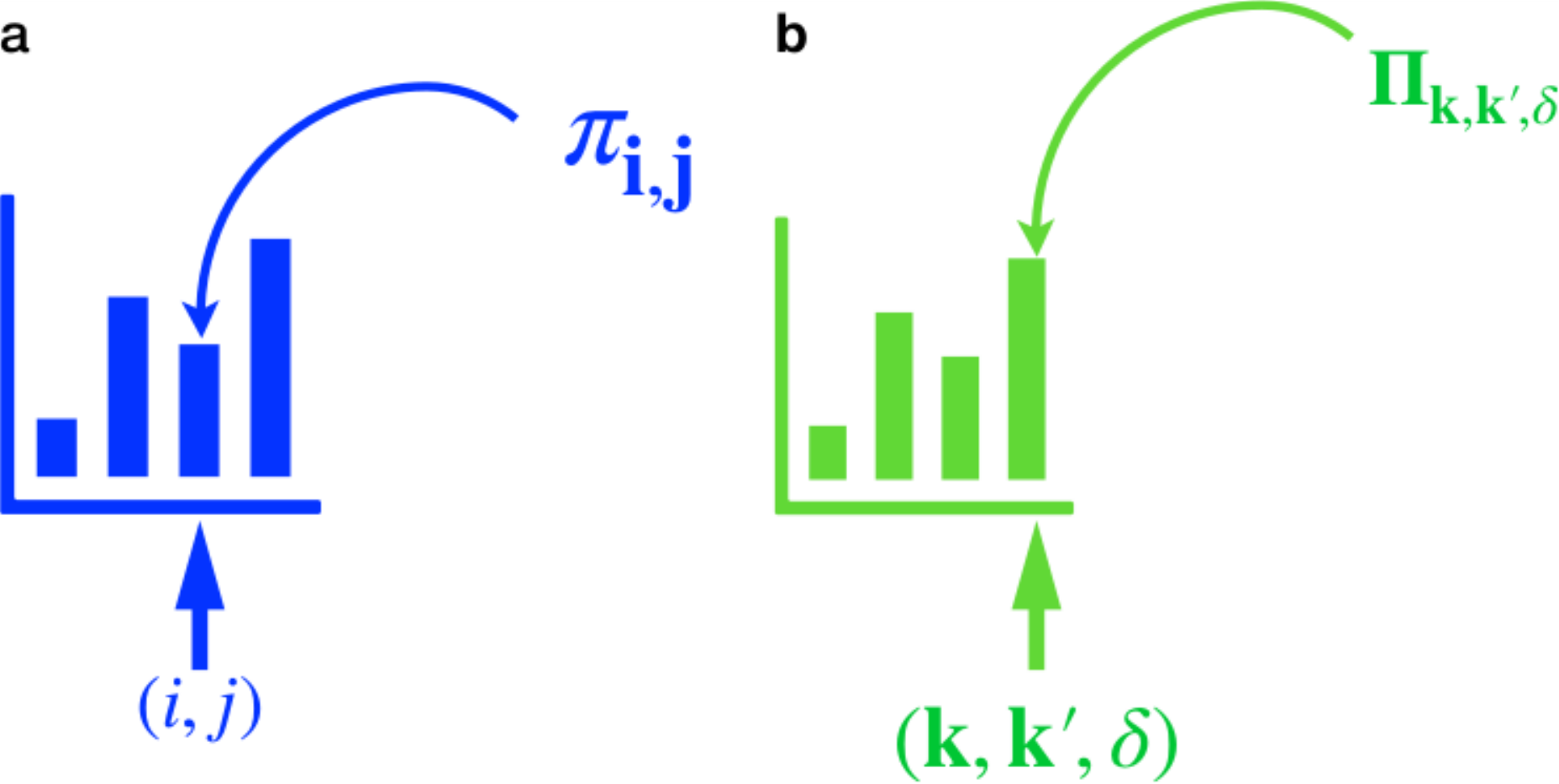}
\caption{{\bf Schematic representation of the classical spatial
    ensembles and their compressed ensemble} (a) The classical spatial ensemble is defined by the probability $\pi_{ij}$ that a link connects node $i$ at one end and node $j$ at the other end.  (b) 
The compressed representation of the ensemble in panel A is defined by
the probability $\Pi_{k,k',\delta}$ that a link connects a node of
degree $k$ at one end and a node of degree $k'$ at distance ${\delta}$
from the first node at the other end.  
}
\label{fig:diagram2}
\end{center}
\end{figure*}

In the following, we apply the  proposed information-theoretical approach to ensembles of spatial networks. We
assume that networks are generated according to different ``sources''
of messages rather than the classical network ensemble with expected degrees, and the
lossy compression of the source of messages consists in replacing link labels  $(i,j)$ with link labels associated in the most general case to $(k,k',\delta)$ where $k$ and $k'$ are the degrees of the linked nodes and $\delta$ is their distance in the underlying space (see Fig. \ref{fig:diagram2}).
The logic behind the formulation of the constrained
maximization problem is still the same as above: we optimize
sources corresponding to similar level of information
for a specified and unchangeable channel.

Our goal is here to show  how the proposed information theory approach   can be used to
predict the most likely distribution of the nodes in space
when pairs of nodes have a given space-dependent
linking probability. 
Our approach reveals that if nodes are
interacting in a network, then interactions induce a natural tendency
of the nodes to be  distributed inhomogeneously in space. The finding
is consistent with the so-called ``blessing of non-uniformity'' of data,
i.e.,  the fact that real-world data 
typically do not obey uniform distributions~\cite{domingos2012few}.
We first consider  spatial networks without any degree constraints,
and then combine spatial and degree-based information in heterogeneous spatial networks.   

\subsection*{Space-dependent linking probability\\}

Let $\delta_{ij}$ be the distance between nodes $i$ and $j$ in some embedding space,
and $\omega(\delta)$ be the distance distribution between all the $N\choose2$ pairs
of nodes, which we also call the correlation function:
the number of pairs of nodes at 
distance $\delta$ is ${N\choose2}\omega(\delta)\simeq
N^2\omega(\delta)/2$.
We define a spatial classical network ensemble by 
imposing the constraint
\bea
\sum_{\vec{\ell}}P(\vec{\ell})\left[\mathbbm{1}(\ell_1 =
  i)+\mathbbm{1}(\ell_2 = j)\right] \, F(\delta_{ij})&=&c
\; ,
\label{eq:constraint_s}
\eea
where $F(\delta)$ is a function of the distance. This constraint can be interpreted as a total "cost" of the links.
Different functions correspond to different ensembles.
For example, in the ensemble with a  cost of the link proportional to the 
their length, this function is  $F(\delta)=\delta$.
If it is $F(\delta)=\ln \delta$, then the cost of a links scales like the
order of magnitude of link lengths.
The maximum entropy principle dictates the maximization
of the functional 
\bea
\hspace{-20mm}{\mathcal G}=S-\mu[\sum_{\vec{\ell}}P(\vec{\ell})-1]-\alpha\left[\sum_{\vec{\ell}}P(\vec{\ell}) \left[\mathbbm{1}(\ell_1 =
  i)+\mathbbm{1}(\ell_2 = j)\right] F(\delta_{ij})-c\right]\; ,
\label{eq:max_entropy_sp}
\eea
leading to 
\bea
\pi_{ij}=\frac{f(\delta_{ij})}{N^2}
\label{pis}
\eea
with  $f(\delta)=g(\delta)/z$, $z=\int d\delta\,\omega(\delta)g(\delta)$,
and $g(\delta)=e^{-\alpha F(\delta)}$.
Therefore if $F(\delta)=\ln \delta$, then the linking probability
decays with the distance as a power law, $g(\delta)=\delta^{-\alpha}$.
If $F(\delta)=\delta$, then this decay is exponential,
$g(\delta)=e^{-\alpha\delta}$.
Fixing the number of links in the network to $L$ as before,
the classical entropy of the ensemble is
\bea
S=-L\sum_{ij}\pi_{ij}\ln \pi_{ij}=\avg{k}N\ln N-\frac{1}{2}\avg{k}N\int d\delta\,\omega(\delta) f(\delta) \ln f(\delta) \; ,
\label{Srho}
\eea
which is the spatial analogue of the classical entropy in 
Eq.~(\ref{eq:classical_entropy}).

We now ask: what is the optimal distribution of nodes in the space at parity
of explicative power of the network model? That is, 
what is the optimal correlation function $\omega^{\star}(\delta)$ for given value of the entropy
$S=S^{\star}$?  To answer this question, we define the entropy $H$ 
of the compressed model in which we consider only the number of ways to distribute $L$ links such that every link connects two nodes at distance $\delta$ with probability  density $\Pi_{\delta}$
\bea
H=-L\int d\delta\,\Pi_{\delta}\ln \Pi_{\delta},\label{eq:spatial_H}
\eea
where $\Pi_{\delta}=L_{\delta}/L$ and $L_{\delta}=L\omega(\delta)f(\delta)$ is the expected number of links between nodes at distance $\delta$ in the classical network ensemble (which clearly satisfy the normalization condition
$\int d\delta\,L_{\delta}=L$).
Our information theory framework shows that the  maximum entropy value of $\omega(\delta)$ is then found by maximizing the functional
\bea
\hspace{-10mm}{\mathcal F}=H -\lambda \left[S -S^{\star}\right]-{\mu}L \left[\int d \delta \, \omega(\delta)f(\delta)-1\right]-{\nu}L \left[\int d \delta \, \omega(\delta)-1\right]\;,\label{eq:spatial_F}
\eea
where $\lambda,\mu$ and $\nu$ are the Lagrange multipliers coupled with the constraints.
The solution reads
\bea
\omega^{\star}(\delta)=e^{-(\mu+1)}e^{-\nu/f(\delta)} f(\delta)^{-(\lambda+1)} \; ,\label{eq:n.vs.f}
\eea
so that $L_{\delta}$ is given by 
\bea
L_{\delta}^{\star}=L\,e^{-(\mu+1)}e^{-\nu/f(\delta)} f(\delta)^{-\lambda}
\; ,
\eea
The Lagrange multipliers are then found as the solutions of the constraints equations.
In Figure $\ref{fig:SH}$(b),  we show the entropy $H$ as a function of
  $S^{\star}$ for a power-law decaying linking probability $f(\delta)=\delta^{-\alpha}/z$.

We now make several important observations. First,
if the space has no boundary, and is isotropic and homogeneous, then
the networks are homogeneous since any two points in the space are
equivalent and the linking probability depends only on the
distance between pairs of points. The degree distribution is thus the
Poisson distribution with the mean equal to the average degree $\avg{k}=2L/N$.
Second, Eq.~(\ref{eq:n.vs.f}) says that the maximum entropy
distribution $\omega(\delta)$ of distances $\delta$ between the nodes 
in the space is uniquely determined by the linking probability $f(\delta)$.
Third, if this probability decays as a power law $f(\delta)=\delta^{-\alpha}/z$,
then the framework describes the natural emergence of power-law pair
correlation functions. Specifically, the solution in Eq.~(\ref{eq:n.vs.f}) decays as a power law at small distances~$\delta$,
while at large distances the decay is exponential due to the finiteness of the system.
If the embedding space is Euclidean 
of dimension $d$, then 
points are scattered in the space according to a fractal distribution. 
Define the node pair density function by
\bea
\rho(\delta)=\frac{\omega^{\star}(\delta)}{\Omega_{\delta}},
\eea
where $\Omega_{\delta}$ is the volume element at distance
$\delta$  from an arbitrary point.
In the $d$-dimensional Euclidean space, $\Omega_{\delta}$ is the volume of the $(d-1)$-dimensional spherical shell, scaling with $\delta$ as 
$\Omega_\delta\propto\delta^{d-1}$.
Therefore for a power-law linking probability $f(\delta)=\delta^{-\alpha}/z$, we get 
  \bea
  \rho(\delta)&\propto&
  \delta^{\beta}e^{-\nu z \delta^\alpha},
  \eea
  where 
$  \beta=(\lambda+1)\alpha-(d-1)$.
Therefore, the embedding in $d$
dimensions is possible only if  $\beta<0$.
Finally, the distribution
of nodes in the space is fractal, and therefore highly nonuniform,
as the uniform distribution would correspond to $\rho(\delta)=const$.

\subsection*{Constraining expected values of node degrees and link costs\\}
As the last example, we consider the classical network ensemble of spatial heterogeneous networks
combining the degree and spatial constraints of Eq.~(\ref{eq:constraint_cm}) and Eq.~(\ref{eq:constraint_s}), respectively.
The probability $\pi_{ij}$ that a random link connects nodes $i$ and $j$ is given by 
\bea
\pi_{ij}
=\frac{\kappa_i\kappa_j}{(\avg{k}N)^2} f(\delta_{ij}),
\eea
where $f(\delta)=e^{-\alpha F(\delta)}/z$, with $\alpha$
the Lagrangian multiplier coupled with the constraint
in Eq.~(\ref{eq:constraint_s}), $z$ the normalization constant
enforcing $\sum_{i,j}\pi_{ij}=1$, and
$\kappa_i$ the hidden variable of node $i$ given
by  $\kappa_i=e^{-\psi_i}\avg{k}N$,  with $\psi_i$ the Lagrangian
multiplier coupled with the constraint in Eq.~(\ref{eq:constraint_cm}).
If there are no correlations between the positions of the nodes in the space and their degrees, then the probability $\pi_{ij}$ can be written as
\bea
\pi_{ij}=\frac{k_ik_j}{(\avg{k}N)^2}f(\delta_{ij}),
\eea
meaning that $\kappa_i=k_i$, so that $\kappa_i$ can be
interpreted as the expected degree $k_i$ of
node $i$.
Using the same approximation as in
Ref.~\cite{krioukov2010hyperbolic} for a power-law decaying function
$f(\delta_{ij})=\delta_{ij}^{-\alpha}/z$, we can write $\pi_{ij}$ as
$\pi_{ij}\propto e^{-r_{ij}}$, where $r_{ij}=\ln \kappa_i+\ln
\kappa_j-\alpha\ln {\delta_{ij}}$ is approximately the
hyperbolic distance between nodes $i$ and $j$ located at radial coordinates
$\ln \kappa_i$ and $\ln \kappa_j$ and at the angular distance proportional to $\delta_{ij}$.
Parameter~$\alpha$ can then be related to the hyperbolic space curvature.
The classical entropy of this ensemble is given by 
\bea
S&=&-L\sum_{ij}\pi_{ij}\ln \pi_{ij}=\avg{k}N\ln
[\avg{k}N]  \nonumber \\
&&\hspace*{-5mm}-\frac{1}{2}\avg{k}N\int d\kappa\int d\kappa'\int d{\delta} \, {\omega}(\kappa,\kappa',\delta)\,\kappa\kappa'f(\delta)\ln [\kappa\kappa'f(\delta)]
\eea
where ${\omega}(\kappa,\kappa',\delta)$ is the density of pairs of nodes with hidden variables $\kappa$ and $\kappa'$ at distance $\delta$.

What is the optimal pair correlation function ${\omega}^{\star}(\kappa,\kappa',\delta)$ for
a fixed value of the classical entropy $S=S^{\star}$?
To answer this question, we maximize the $H$-entropy of the compressed model 
 \bea
H=-L\int d\kappa \int d{\kappa'}\int d\delta\,
\Pi_{\kappa,\kappa',\delta}\ln \Pi_{\kappa,\kappa',\delta} \; , \eea
where $\Pi_{\kappa,\kappa',\delta}=L_{\kappa,\kappa',\delta}/L$ is the probability density that a link connected two nodes of with hidden variables $\kappa$ and $\kappa'$ and at distance $\delta$.
Note that $L_{\kappa,\kappa',\delta}=L\omega({\kappa,\kappa',\delta})\kappa \kappa' f\left(\delta\right)$ indicates  the expected number of links between pairs of
nodes with hidden variables $\kappa$ and $\kappa'$ at distance $\delta$ in the classical network ensemble.
The maximization of $H$ under the constraints $S=S^{\star}$, the normalization of $L_{\kappa,\kappa',\delta}$, $\int d\kappa \int d\kappa'\int d\delta \, L_{\kappa,\kappa',\delta}=L$, and the normalization of $\omega(\kappa,\kappa',\delta)$, $\int d\kappa \int d\kappa'\int d\delta \, \omega({\kappa,\kappa',\delta})=1$, yields the answer
\bea
\omega^{\star}(\kappa,\kappa',\delta)=e^{-(\mu+1)}\exp\{-\nu/[\kappa \kappa'
f\left(\delta\right)]\} [\kappa \kappa'
f\left(\delta\right)]^{-(\lambda+1)},\label{omegaz}
\eea
where $\lambda,\mu$ and $\nu$ are the Lagrange multipliers coupled with the $S=S^\star$ constraint,
the normalization of $L_{\kappa,\kappa',\delta}$, and the 
normalization of ${\omega}(\kappa,\kappa',\delta)$, respectively.
Observe that the pair correlation function ${\omega}(\kappa,\kappa',\delta)$ depends
on its arguments only via $w=\kappa\kappa' f\left(\delta\right)$, and for small values of $w$ it decays as a
power-law function of $w$. If $f(\delta)=\delta^{-\alpha}/z$, then
$\omega(\kappa,\kappa',\delta)$ can be also written in terms of the  
approximate hyperbolic distance $r=\ln w = \ln \kappa+\ln \kappa'-\alpha\ln {\delta}$ as
\bea
\omega^{\star}(r)=e^{-(\mu+1)}\exp[-{(\lambda+1)}r-\nu e^{-r}] \; .
\eea
As in the homogeneous spatial case, here we also observe that
the optimal distribution of nodes in the space is
not uniform.

\begin{figure*}[!htbp]
\begin{center}
\includegraphics[width=0.95\textwidth]{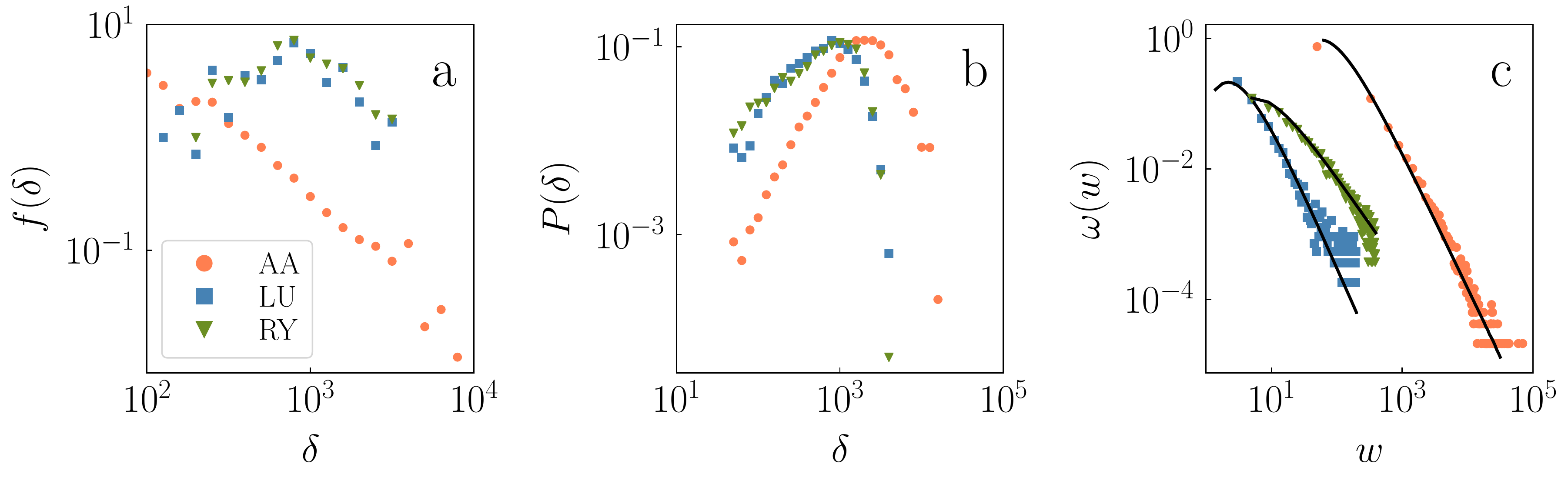}
\caption{{\bf The application of the information-theoretical framework to real-world airport networks.} 
The networks correspond to the flights operated by American Airlines
(AA)
during January-April 2018 between US airports~\cite{bureau}, by
Lufthansa (LU) and Ryanair (RY) during year 2011 between European airports ~\cite{cardillo}.
For each air carrier,
a separate air transportation network is built, in which nodes are
airports and two airports are connected if at least one flight
between the two airports is present in the data. Using the
network topology and the geographic locations of the airports, 
the empirical linking probability $f(\delta)$ (panel~(a))
and the density distribution $P(\delta)=\int d\kappa \int d\kappa'
\omega(\kappa,\kappa',\delta)$ 
(panel (b)) are computed for the three
networks, where distance~$\delta$ is geographic and is measured in kilometers.
Panel~(c) shows the pair correlation functions
$\omega(\kappa,\kappa',\delta)=\omega(w)$, where $w=\kappa\kappa'f(\delta)$, for the three networks.
Points represent empirical densities, while the full lines are
theoretical predictions according to Eq.~(\ref{omegaz}). Values of the
Lagrange multipliers are:  $\lambda = 1.2$ and $\nu = 120$ for AA,
$\lambda = 1.3$ and $\nu =5 $ for LU, and $\lambda =0.45$ and $\nu = 8$ for RY.}
\label{fig:airports}
\end{center}
\end{figure*}

\subsection*{Real-world networks\\}

In Figure~$\ref{fig:airports}$ we apply the considered information-theoretic framework 
to real-world air transportation networks, 
in which nodes are airports and edges between pairs of nodes indicate the
existence of at a least one flight connecting the two airports. Specifically, we
consider three networks corresponding to 
flights operated in different geographic areas by three air carriers.
The distances $\delta_{ij}$ between airports $i$ and $j$ are their geographic distances.
The linking probability $f(\delta)$ is computed from the data as the empirical
connection probability, and the hidden variables $\kappa_i$ are set to the
actual degrees of the airports in the networks.
We note that the empirical connection probabilities $f(\delta)$
decay as power laws, 
and that  the pair correlation functions $\omega(w)$ are well described by 
Eq.~(\ref{omegaz}).

\section*{Discussion}
In summary, this work illustrates a classical information-theoretical
approach to the characterization of random networks. This framework is based on a tradeoff between the entropy of the network ensemble and the entropy of its compressed representation. According to our
theory, network inhomogeneities in the distribution of node degrees 
and/or node position in space both emerge from the 
general principle of maximizing randomness at parity of explicative
power.
The framework provides theoretical foundations for a series of
models often encountered in network science, and
can likely be extended to generalized network
models such as multilayer networks and simplicial
complexes~\cite{bianconi2013statistical,courtney2016generalized} or to information theory approaches based on the network spectrum  \cite{de2016spectral}.
In applications to real-world networks, the framework provides a theoretical    explanation of the  nontrivial 
 inhomogeneities  that are an ubiquitous features of real-world complex systems.

\section*{Acknowledgements}
 F. R. acknowledges support from NSF Grant No. CMMI-1552487 and ARO Grant No. W911NF-16-1- 0104.
D. K. acknowledges support from ARO  Grant No.  W911NF-16-1-0391 and No. W911NF-17-1-0491 and NSF Grant No. IIS-1741355.
\section*{Data availability.}
The networks considered in this study are generated from data
corresponding to flights operated by American Airlines
(AA) during January-April 2018 between US airports~\cite{bureau}, and by
Lufthansa (LU) and Ryanair (RY) during year 2011 between European
airports ~\cite{cardillo}. Geographical coordinates of the
airports have been obtained from \url{https://openflights.org/data.html}.
AA data are available upon request.
LU and RY data can be downloaded from the
repository Air Transportation Multiplex at \url{http://complex.unizar.es/~atnmultiplex/}.
\section*{Code availability.}
Code used to generate Figures~2 and~4 is available upon request.
\section*{References}
\bibliographystyle{unsrt}
\bibliography{bibliography}

\appendix

\section{The classical network ensemble}
We consider a
classical network ensemble
defining the
probability of a network $G=(V,E)$ of $|V|=N$ nodes and $|E|=L$
links. In this ensemble, a network $G$ is described by an edge list $\left\{\vec{\ell}^{[n]}\right\}$ with $n\in\{1,2\ldots, L\}$ where each link $\vec{\ell}^{[n]}$ is described by an ordered pair of node labels  $\vec{\ell}^{[n]}=\left(\ell_1^{[n]},\ell_2^{[n]}\right)$, i.e.,  $\ell_1^{[n]}$  indicates the label of the node $i\in\{1,2,\ldots, N\}$ attached to the first end of the link $n$ and similarly $\ell_2^{[n]}$ indicates the label of the node attached to the second end of the link $n$.

Every link variable $\vec{\ell^{[n]}}$ can assume values of the type $(i^{[n]},j^{[n]})$ with $i,j\in\{1,2,\ldots, N\}$.
In the classical network ensemble, every link is independently
distributed, thus we associate to each  network
(edge list)  $\{\vec{\ell}^{n}\}$ a probability 
\bea
\mathbb{P}\left(G\right)=\prod_{n=1}^L P\left(\vec{\ell}^{[n]}\right)
\eea
where $P(\vec{\ell^{[n]}})$ is the probability that the $n$-th link is connected to the pair of nodes  $(\ell^{[n]}_1,\ell^{[n]}_2)$.
The entropy of this ensemble is given by 
\bea
S(G)=-L\sum_{\vec{\ell}}P(\vec{\ell})\ln P(\vec{\ell}).
\eea
Note that alternatively we could define the network ensemble as 
given by a  set of $L$ undistinguishable links defined as unordered
pairs of node labels $\vec{\ell}$. In that case,
by following similar mathematical steps as those used to treat the Gibbs paradox~\cite{Huang} in statistical mechanics,
the entropy would only differ by a constant term, i.e., 
\bea
S^{\left[\mbox{undis}\right]}(G)&=&-\sum_{\{\vec{\ell}\}}\mathbb{P}(\{\vec{\ell}\})\ln \mathbb{P}(\{\vec{\ell}\})\nonumber \\&=&-L\sum_{\vec{\ell}}P(\vec{\ell})\ln P(\vec{\ell})-\ln (L! 2^L).
\eea
The above entropy might be preferred to the entropy $S$ associated to
distinguishable links. However, the $S$ and
$S^{\left[\mbox{undis}\right]}$ entropies
differ only by a global term that depends on the total number of links
only, thus making $S$ and $S^{\left[\mbox{undis}\right]}$ equivalent
for the purpose of our mathematical framework.
We further note that the classical network ensemble is fully described
by the link ensemble.
The link ensemble is a triple $(\vec{\ell},{\mathcal A}_{\vec{\ell}}, {\mathcal P}_{\vec{\ell}})$ where $\vec{\ell}$ indicates the value associated of the random variable associated to an arbitrary link of the network, ${\mathcal A}_{\ell}=\{(i,j)| i,j\in \{1,2,\ldots,N\}\}$ indicates the set of all distinct possible values that the link random variable can assume, and $ {\mathcal P}_{\ell}=\{\pi_{ij},i,j\in \{1,2,\ldots,N\}\}$ indicates the set of probabilities 
\bea
\mathbb{P}(\vec{\ell}=(i,j))=\pi_{i,j} \; .
\eea
 Here, we consider maximum entropy classical network ensembles where
 the probabilities $\pi_{ij}$
 only depend on some hidden variables ${\bf x_{ij}}$ associated to the
 link, i.e.,  where
 \bea
 \pi_{ij}=\pi_{\bf x_{ij}}.
 \eea
 Alternatively, we could say that $\pi_{i,j}$ is the probability that $\vec{\ell}=(i,j)$ given that the two nodes are characterized by the hidden variables ${\bf x}_{ij}$ assigned a priori to each pair of nodes of the network. 
 For instance, if we consider the classical ensemble in which we constraint the expected degree sequence, we will have 
 \bea
 \pi_{i,j}=\pi_{k_i,k_j}.
 \eea 
 while in the spatial network we will have 
 \bea
 \pi_{i,j}=\pi_{\kappa_i,\kappa_j,\delta_{ij}}.
 \eea
 Thus, the entropy $S(G)$ of the classical network ensemble is given by  
 \bea
 S(G)=-L\sum_{i,j}\pi_{ij}\ln \pi_{ij}=-L\sum_{{\bf x}}N^2P_{\mathcal V}({\bf x})\pi_{\bf{x}}\ln \pi_{\bf x},
 \label{Scond}
 \eea
 where 
 \bea
 P_{\mathcal V} ({\bf x})=\frac{1}{N^2}\sum_{i,j}\delta({\bf
   x}_{ij},{\bf{x}}) 
 \eea
 indicates the probability that a random pair of nodes has hidden variable ${\bf x}_{ij}={\bf{x}}$.
 Since the network ensemble is constructed given the distribution of
 hidden variables, the entropy $S(G)$ can be interpreted as a
 conditional entropy of the network given the distribution of hidden
 variables as the rightmost term of Eq.(\ref{Scond}) reveals. 
 \section{The channel that compresses information}
 It follows that a classical network ensemble can be considered as a
 source of $L$ messages. Each message is a link $\vec{\ell}$ carrying information on the node labels of the two linked nodes.
We assume that the information is compressed by a channel $Q$, characterized by an input $\vec{\ell}$ taking values in ${\mathcal A}_{\vec{\ell}}$ and an output ${\bf x}(\vec{\ell})$ indicating the  hidden variables associated to the link
\bea
{\bf x}(\vec{\ell})={\bf x}_{ij}.
\eea
The channel $Q$ is a lossy compression channel that is erasing
information about the identity of the nodes, and retaining only the
value of their
hidden variables. The output of the channel $Q$ is the ensemble
$\{{\bf x(\vec{\ell})},{\mathcal A}_{\bf x},{\mathcal P}_{\bf x}\}$,
where the random variables associated to
each link  are given by the hidden variables of the linked nodes ${\bf x}(\ell)$. 
${\mathcal A}_{\bf x}$  is the set of  all possible values
that the hidden variables of a link
can take, and ${\mathcal P}_{\bf x}$ is the set of all probabilities 
\bea
\Pi_{\bar{\bf x}}=\mathbb{P}({\bf x}(\vec{\ell})=\bar{\bf x})=\sum_{i<j}\pi_{ij}\delta({\bf x}(\vec{\ell},\bar{\bf x}))=\pi_{\bf \bar{x}}N^2P_{\mathcal V}(\bar{\bf x}).
\eea
For instance, if the hidden variables are exclusively the expected degrees of the nodes, we have 
\bea
\Pi_{k,k^{\prime}}&=&\mathbb{P}({\bf x}(\vec{\ell})=(k,k^{\prime}))=\sum_{i<j}\pi_{ij}\delta(k_i,k)\delta(k_j,k^{\prime})\nonumber \\&=&\pi_{k,k^{\prime}}N^2P(k)P(k^{\prime}).
\eea
If instead we are considering a spatial network with hidden variables
${\bf x}=(\kappa,\kappa^{\prime},\delta)$,  we have
\bea
\Pi_{k,k^{\prime},\delta}&=&\mathbb{P}({\bf
  x}(\vec{\ell})=(k,k^{\prime},\delta))=\sum_{i<j}\pi_{ij}\delta(k_i,k)\delta(k_j,k^{\prime})\delta(\delta_{ij},\delta)\nonumber
\\&=&N^2\pi_{k,k^{\prime,\delta}}\omega(\kappa,\kappa^{\prime},\delta) ,
\eea
where here we have adopted the notation of the main text.
The output message defines a compressed network ensemble  of networks  having  entropy
\bea
H=-L\sum_{\bf x}\Pi_{\bf x}\ln \Pi_{\bf x}=-L\sum_{\bf x}N^2\pi_{\bf
  x}P_{{\mathcal V}}({\bf x})\ln \left(N^2\pi_{\bf x} P_{{\mathcal
      V}}({\bf x})\right) .
\eea
Interestingly, we have that the entropy $H$ is equal to the mutual information between the input message and the output messages of the channel $Q$ multiplied by $L$, i.e.,
\bea
H=L \sum_{\vec{\ell},{\bf x}}P(\vec{\ell},{\bf x})\ln \left(\frac{P(\vec{\ell},{\bf x})}{P(\vec{\ell})P({\bf x})}\right).
\eea
This fact follows immediately from the observation that the joint distribution $ P(\vec{\ell},{\bf x})$ is simply given by 
 \bea
 P(\vec{\ell},{\bf x})=P(\vec{\ell})\delta({\bf x}_{ij},{\bf x}) ,
 \eea
 i.e., the value of ${\bf x}(\vec{\ell})$ is uniquely determined by $\vec{\ell}$ and the relation 
 \bea
 P({\bf x})=\sum_{\vec{\ell}}P(\vec{\ell},{\bf x})=\Pi_{\bf x}.
 \eea

\section{Optimal hidden variable distribution}

Our framework aiming at finding the optimal distribution of hidden
variables $P_{\mathcal V}^{\star}({\bf x})$ consists in maximizing $H$
for a fixed value of $S=S^{\star}$.
In particular,  we define the
optimal hidden variable distribution
$P_{{\mathcal V}}^{\star}({\bf x})$
as the solution of the optimization problem
\bea
P_{{\mathcal V}}^{\star}({\bf x})=\arg\max_{{P_{\mathcal V}}({\bf x})} [H-\lambda S].
\label{principle}
\eea 
under the constraints the  network contains a given number 
of nodes and links,  and that $S=S^{\star}$.
Since $H$ is proportional to the mutual information of the channel
$Q$,
the maximum of $H$ given $S=S^{\star}$ consists in the  capacity of
the channel under the constraint that the network ensemble has entropy $S=S^{\star}$. 
Interestingly, our optimal hidden variable distribution can be seen as
a parallel of the {\it optimal input distribution}
\cite{mackay2003information} of a channel, with the difference that
consider a network model where $\pi_{\bf x}$ fixed 
and we optimize only the distribution of the hidden variables ${P_{\mathcal V}}({\bf x})$.
\end{document}